\documentstyle[preprint,tighten,pra,aps]{revtex}
\begin{document}
\draft
\title{Kinks in the Presence of Rapidly Varying Perturbations}

\author{Yuri S. Kivshar}
\address{Optical Sciences Centre Australian National University\\
Australian Capital Territory 0200 Canberra, Australia}

\author{Niels Gr{\o}nbech-Jensen}
\address{Theoretical Division, Los Alamos National Laboratory\\
Los Alamos, NM 87545}

\author{Robert D. Parmentier}
\address{Dipartimento di Fisica, Universit\`{a} di Salerno\\
I-84081 Baronissi (SA), Italy}

\maketitle

\begin{abstract}
Dynamics of sine-Gordon kinks in the presence of rapidly varying periodic
perturbations of different physical origins is described analytically and
numerically.  The analytical approach is based on asymptotic expansions,
and it allows to derive, in a rigorous way, an effective nonlinear equation
for the slowly varying field component in any order of the asymptotic
procedure as expansions in the small parameter $\omega^{-1}$, $\omega$
being the frequency of the rapidly varying ac driving force.  Three
physically important examples of such a dynamics, {\em i.e.}, kinks driven
by a direct or parametric ac force, and kinks on rotating and oscillating
background, are analysed in detail.  It is shown that in the main order of
the asymptotic procedure the effective equation for the slowly varying
field component is {\em a renormalized sine-Gordon equation} in the case of
the direct driving force or rotating (but phase-locked to an external ac
force) background, and it is {\em the double sine-Gordon equation} for the
parametric driving force.  The properties of the kinks described by the
renormalized nonlinear equations are analysed, and it is demonstrated
analytically and numerically which kinds of physical phenomena may be
expected in dealing with the renormalized, rather than the unrenormalized,
nonlinear dynamics. In particular, we predict several qualitatively new
effects which include, {\em e.g.}, the perturbation-induced internal
oscillations of the $2\pi-$kink in a parametrically driven sine-Gordon
model, and generation of kink motion by a pure ac driving force on a
rotating background.
\end{abstract}

\pacs{PACS numbers: 03.40.Kf, 74.40.+b, 74.50.+r, 85.25.Cp}

\section{Introduction}

As is well known, the effect of rapidly varying perturbations on the
dynamics of nonlinear systems may lead to a drastic change of the system
behavior in the sense of the {\em averaged} dynamics. In particular,
large-amplitude {\em parametric} perturbations may give rise to {\em a
stabilization} of certain types of dynamical regimes.  A typical and famous
example is a stabilization of a reverse pendulum by parametric forced
oscillations of its pivot \cite{1} (see also the recent paper \cite{N} and
references therein), and a similar effect may be also achieved by applying
a direct ac driving force of large amplitude \cite{2}.  Such a dynamical
stabilization has its analog in nonlinear systems with distributed
parameters supporting, in particular, novel types of kink solitons
\cite{3,4,5,6}.  However, the method which is usually used to derive an
averaged equation describing the system dynamics in the presence of rapidly
varying perturbations is not rigorous and, as a matter of fact, it is not
well justified.  Such a method, even being very clear from the physical
point of view, uses a splitting of slow and fast variables and subsequent
averaging which is based, in fact, on solutions of a linearized equation
for fast variations where the slowly varying coefficients are assumed to be
constant \cite{1,3,4}.  The procedure of such a linearization assumes that
the amplitude of the forced (rapidly varying) oscillations is small, and
this is certainly valid for parametrically forced oscillations far from the
parametric resonance.  For direct ac perturbations, the forced oscillations
may become large.  To describe the dynamics in an approximate way, the
so-called ``rotating-wave approximation'' was used without detailed
mathematical justification \cite{5,6,NL}.  It is necessary to note that the
derivation of an effective averaged equation for the slowly varying field
component is an important (and nontrivial) step of the analysis of such
systems, and in many of the cases the corresponding equation determines the
leading physical effects observed in the presence of rapidly varying
perturbations. In all the cases it is necessary to justify the averaging
procedure as well as to estimate the influence of the higher-order
contributions. Unfortunately, the latter are beyond the usual averaging
methods.  However, as we show in the present paper, a rigorous analysis of
the effect of rapidly varying periodic perturbations on nonlinear dynamics
of ac driven damped systems can be performed in a straightforward way to
describe the averaged dynamics with any accuracy.

The purpose of this paper is to present the basic steps of the method
mentioned above and, selecting the sine-Gordon (SG) model as a particular
example, to describe the dynamics of kinks in the presence of rapidly
varying driving forces of very different physical origins.  Considering the
external ac driving force to be rapidly oscillating, we apply an asymptotic
procedure based on a Fourier series where the coefficients are assumed to
be slowly varying functions on the time scale $\omega^{-1}$, $\omega$ being
the frequency of the rapidly varying ac force which is assumed to be large.
The basic idea to split fast and slow variables is not new, and the
well-known example is, as mentioned, a stabilization of the reverse pendulum
by oscillations of its suspension point.  However, our analytical method to
derive an effective equation for the slowly varying field component is
novel, and this method allows us to calculate, in a self-consistent way,
all the corrections using solely asymptotic expansions rather than direct
averaging in fast oscillations. It is clear that the applicability of the
method itself is much wider than the particular examples covered by the
present paper.

The paper is organized as follows. In Sec. II we consider the case of a
direct ac driving force demonstrating the basic steps of our analytical
approach in detail. The main result of such an analysis is the so-called
``averaged'' equation, {\em i.e.}, that describing slowly varying system
dynamics.  In the case of the direct driving force this equation is shown
to be a renormalized SG equation. Section III presents the case of a
parametric driving force where the final equation describing the slowly
varying dynamics is the double SG equation which, as we show, may display
new features in the averaged kink dynamics, {\em e.g.}, oscillations of the
excited internal mode of the kink, which is absent in the standard SG
model. The case of kink stabilization on a rotating background by applying
a rapidly oscillating ac force is discussed in Sec. IV. There we analyse an
effect of an induced dc force on the kink motion, as well as showing
numerically that the main conclusions of our analysis may be easily
extended to cover multi-soliton dynamics.  Finally, Sec. V concludes the
paper.

\section{Direct driving force}

\subsection{Asymptotic expansions}

As the first example, we consider the case of the direct driving force in
the SG model when the system dynamics is described by the driven damped SG
equation for the field variable $\phi(x,t)$,
\begin{equation}
\label{1}
\frac{\partial^2 \phi}{\partial t^2} - \frac{\partial^2 \phi}{\partial x^2}
+ \sin \phi = f - \gamma \frac{\partial \phi}{\partial t} + \epsilon \cos
(\omega t),
\end{equation}
where $f$ is a constant contribution of the driving force, $\gamma$ is the
damping coefficient, and the amplitude $\epsilon$ of the driving force may
be large (in fact, up to the values of order of $\omega^2$).  The standard
physical application of the model (\ref{1}) is  to describe the fluxon
dynamics in long Josephson junctions (see, {\em e.g.}, \cite{McS}), so that
$f$ and $\epsilon \cos (\omega t)$ are the constant and varying components
of the bias current applied to the junction. In the subsequent analysis we
consider the direct driving force ($\sim \epsilon$) as {\em rapidly
oscillating}, {\em i.e.}, the frequency $\omega$ is assumed to be large in
comparison with the frequency gap $(=1)$ of the linear spectrum band.  Our
purpose is to derive an ``averaged'' nonlinear equation to describe the
{\em slowly varying} dynamics of the SG field.

In order to derive an averaged equation of motion, we note first that in
the case of very different time scales the SG field  $\phi$ may be
decomposed into a sum of slowly and rapidly varying parts, {\em i.e.},
\begin{equation}
\label{2}
\phi = \Phi + \zeta.
\end{equation}
The function $\zeta$ stands for fast oscillations around the slowly varying
envelope function $\Phi$, and the mean value of $\zeta$ during an
oscillation period is assumed to be zero so that $<\phi> = \Phi$.  Our goal
is to derive an effective equation for the function $\Phi$. The standard
way to do that is to substitute Eq. (\ref{2}) into Eq. (\ref{1}) and to
split Eq. (\ref{1}) into two equations for slow and fast variables, making
an averaging to obtain the equation for the slowly varying field component
(see, e.g., \cite{3,4}). However, such an approach must be properly
justified for the case when the fast oscillations {\em are not small} as it
is for the direct driving force considered here, and in a similar problem
it was proposed \cite{5} to use the so-called ``rotating wave
approximation'' to find the rapidly oscillating field component.  All these
approaches, although quite satisfactory for the first-order approximation
(see, e.g., \cite{3,5,6}), do not allow to make the next-order expansions
to calculate higher-order corrections, and thus they cannot be rigorously
justified.  Nevertheless, as we show in the present paper, a rigorous
approach may indeed be proposed to obtain an effective equation for the
slowly varying field component $\Phi$ {\em with any accuracy}.

The basis of our asymptotic procedure is a Fourier series expansion with
slowly varying coefficients. We look for rapidly oscillating component
$\zeta$ in the form
\begin{equation}
\label{5}
\zeta = A \cos(\omega t) + B \sin(\omega t) + C \cos(2 \omega t) + D \sin(2
\omega t) + \ldots \; ,
\end{equation}
where the coefficients $A$, $B, \ldots$ are assumed to be slowly varying on
the time scale $\sim \omega^{-1}$.  Substituting the expressions (\ref{2}),
(\ref{5}) into Eq.  (\ref{1}), we note that the effective coupling between
different harmonics of the expansion (\ref{2}), (\ref{5}) is produced by
the nonlinear term $\sin \phi$, which generates the following Fourier
expansion,
\begin{eqnarray}
\sin \phi = \sin \Phi \left[ \alpha_0 + \alpha_1 \cos (\omega t) + \alpha_2
\sin (\omega t) + \alpha_3 \cos (2\omega t) + \alpha_4 \sin (2\omega t) +
\ldots \right] \nonumber \\ +  \cos \Phi \left[ \beta_0 + \beta_1 \cos
(\omega t) + \beta_2 \sin (\omega t) + \beta_3 \cos (2\omega t) + \beta_4
\sin (2\omega t) + \ldots  \right],
\end{eqnarray}
where
\begin{eqnarray}
\alpha_0 = J_0(A) \left( 1 -\frac{1}{4} B^2\right) + \frac{1}{4} B^2 J_2 (A)
+ \ldots , \;\;\; \alpha_1 = - C J_1 (A) + \ldots , \nonumber \\ \alpha_2 =
-D J_1(A) + \ldots , \;\;\; \alpha_3 = \frac{1}{4} B^2 J_0(A) + \ldots, \;\;
\; \alpha_4 = - BJ_1(A) + \ldots ,
\end{eqnarray}
\begin{eqnarray}
\beta_0 = - CJ_2(A) + \ldots, \;\;\; \beta_1 = 2J_1(A) + \ldots, \;\;\;
\beta_2 = B J_0(A) + BC J_1(A) + \ldots, \nonumber \\ \beta_3 = C J_0(A) +
\ldots, \;\;\; \beta_4 = D J_0(A) + \ldots,
\end{eqnarray}
and $J_0$, $J_1$, {\em etc.}, are Bessel functions. Collecting now the
coefficients in front of the different harmonics, we obtain an infinite set
of coupled nonlinear equations,
\begin{eqnarray}
\label{4}
\frac{\partial^2 \Phi}{\partial t^2} - \frac{\partial^2 \Phi}{\partial x^2}
+ \sin \Phi \left[ J_0(A) \left(1 - \frac{1}{4} B^2\right) + \frac{1}{4}
B^2 J_2(A) + \ldots \right] \nonumber\\ + \cos \Phi \left[ - C J_2(A) +
\ldots \right] = f - \gamma \frac{\partial \Phi}{\partial t},
\end{eqnarray}
\begin{eqnarray}
\label{6}
\left(- \omega^2 A + 2 \omega \frac{\partial B}{\partial t} +
\frac{\partial ^2 A}{\partial t^2}\right) - \frac{\partial^2 A}{\partial
x^2} + \cos \Phi [ 2 J_1(A) + \ldots ] \nonumber\\ + \sin \Phi [ - C J_1(A)
+ \ldots ] + \gamma \left( \frac{\partial A}{\partial t} + \omega B\right)
= \epsilon,
\end{eqnarray}
\begin{eqnarray}
\label{7}
\left( -\omega^2 B - 2 \omega \frac{\partial A}{\partial t} +
\frac{\partial^2 B}{\partial t^2} \right) - \frac{\partial^2 B}{\partial
x^2} + \cos \Phi [ B J_0(A) + B C J_1(A) + \ldots ] \nonumber\\ + \sin \Phi
[ - D J_1(A) + \ldots ] + \gamma \left( \frac{\partial B}{\partial t} -
\omega A \right) = 0,
\end{eqnarray}
\begin{eqnarray}
\label{8}
\left( - 4\omega^2 C + 4\omega \frac{\partial D}{\partial t} +
\frac{\partial^2 C}{\partial t^2} \right) - \frac{\partial^2 C}{\partial
x^2} + \cos \Phi [ C J_0(A) + \ldots ] \nonumber \\ +\sin \Phi \left[
\frac{1}{4} B^2 J_0(A) + \ldots \right] + \gamma \left(\frac{\partial
C}{\partial t} + 2\omega D\right) = 0,
\end{eqnarray}
\begin{eqnarray}
\label{9}
\left(- 4\omega^2 D - 4\omega \frac{\partial C}{\partial t} +
\frac{\partial^2 D}{\partial t^2}\right) - \frac{\partial^2 D}{\partial
x^2} + \cos \Phi [ D J_0(A) + \ldots ] \nonumber \\ + \sin \Phi [ - B J_1(A)
+ \ldots ] + \gamma \left( \frac{\partial D}{\partial t} - 2\omega C \right)
= 0,
\end{eqnarray}
and similar equations for the coefficients in front of the higher-order
harmonics. To proceed further, we note that different terms in Eqs.
(\ref{6}) to (\ref{9}) are not equivalent provided $\omega$ is a large
parameter. Indeed, if we assume the amplitude $\epsilon$ large as well
(otherwise, the dynamics of the system is rather trivial because the effect
of small-amplitude but rapidly oscillating force is negligible), let us say
up to the order of $\omega^2$, the large term $-\omega^2 A$ in Eq. (\ref{6})
may be compensated only by the term $\epsilon$ from the right-hand side of
Eq. (\ref{6}). Thus, assuming $\epsilon \sim \omega^2$ we find the first
term of the asymptotic expansion $A \approx - \epsilon/\omega^2$. On the
other hand, the right-hand side of Eq. (\ref{7}) is zero, and the large
term $-\omega^2 B$ may be compensated only by a contribution from the other
terms $\sim A$, thus giving the first term of the expansion for the
coefficient $B$, {\em viz.}, $B \approx \gamma \epsilon/\omega^3$. Such a
simple reasoning may be effectively applied to other coefficients as well
as to other corrections of the asymptotic expansion. As a matter of fact,
to generalize and simplify the procedure of calculation of the expansion
coefficients, we look for the coefficients $A, B, \ldots$ in the form of
the power series in the small parameter $\omega^{-1}$ as follows
\begin{equation}
\label{11}
A=a_{1}+\frac{a_{2}}{\omega^2}+ \ldots, \;\; B=\frac{b_{1}}{\omega}+\frac{b_
{2}}{\omega^3} + \ldots, \;\; C = \frac{c_{1}}{\omega^4} +
\frac{c_2}{\omega^6} + \ldots, \;\; D=\frac{d_{1}}{\omega^3}+
\frac{d_2}{\omega^5} + \ldots \;.
\end{equation}
Substituting Eq. (\ref{11}) into Eqs. (\ref{6}) to (\ref{9}) and equating
the terms of the same orders in the small parameter $\omega^{-1}$, we find
\begin{equation}
\label{12}
a_{1}= - \frac{\epsilon}{\omega^2} \equiv - \delta, \end{equation}
\begin{equation} \label{13} a_{2}= \gamma b_1 + 2 \cos \Phi \, J_1(a_1),
\end{equation}
\begin{equation}
\label{14}
b_{1}= -  \gamma a_1,
\end{equation}
\begin{equation}
\label{15}
b_{2}= - 2 \frac{\partial a_2}{\partial t} - \gamma a_2  + b_1 \cos \Phi
[J_0 (a_1) + 2 J_2(a_1)],
\end{equation}
\begin{equation}
\label{16}
c_{1}=  \frac{1}{4} \left[ 4 \frac{\partial d_1}{\partial t} + \frac{1}{4}
b_1^2 J_0(a_1) \sin \Phi + \frac{1}{2} b_1^2 J_2(a_1) + 2\gamma d_1 \right],
\end{equation}
\begin{equation}
\label{16a}
d_{1}= - \frac{1}{4} b_1 J_1(a_1) \sin \Phi,
\end{equation}
and so on. In Eq. (\ref{12}) the parameter $\delta = \epsilon/\omega^2$ is
assumed to be of order of ${\cal O}(1)$, but all the results are valid also
for the case $\delta \ll 1$.  The expansions (\ref{11}) allow to find the
coefficients $A, B, \ldots$ in each order of $\omega^{-1}$, and all the
corrections are determined by {\em algebraic} relations rather than
additional differential equations.  For example, $a_{2}$ is determined by
Eq. (\ref{13}) through $b_{1}$ which, in its turn, may be found from Eq.
(\ref{14}) as a function of $a_{1}$, {\em i.e.}, through the slowly varying
part $\Phi$, and so on. This statement is valid for all coefficients of the
asymptotic expansion: The coefficients are found through {\em algebraic}
relations involving lower-order terms of the asymptotic expansions and
their derivatives, and one does not need to find solutions of additional
differential equations.

\subsection{Renormalized equation}

Applying the expansions (\ref{11}) to Eq. (\ref{4}), we may find the
equation for the slowly varying field component $\Phi$ with any accuracy in
the small parameter $\omega^{-1}$, {\em e.g.},
\begin{eqnarray}
\label{17}
\frac{\partial^2 \Phi}{\partial t^2} - \frac{\partial^2 \Phi}{\partial x^2}
+ \sin \Phi \left[ J_0(a_1) + \frac{1}{\omega^2} \left( - \frac{1}{4} b_1^2
J_0(a_1) - a_2 J_1(a_1) + \frac{1}{4} b_1^2 J_2(a_1) \right) + \ldots
\right] \nonumber\\ + \cos \Phi \left[ -\frac{c_1}{\omega^4} J_2(a_1) +
\ldots \right] = f - \gamma \frac{\partial \Phi}{\partial t}.
\end{eqnarray}
Thus, from the asymptotic procedure described above it is quite obvious how
to calculate the corrections of the first, second, and subsequent orders
and to find the averaged equation with any required accuracy.

In the first-order approximation in $\omega^{-1}$ only the term $ J_0(a_1)
\sin \Phi$ contributes, so that Eq. (\ref{17}) yields
\begin{equation}
\label{18}
\frac{\partial^2 \Phi}{\partial t^2} - \frac{\partial^2 \Phi}{\partial x^2}
+ J_0 \left(\frac{\epsilon}{\omega^2}\right) \sin \Phi = f - \gamma
\frac{\partial \Phi}{\partial t}.
\end{equation}

Equation (\ref{18}) takes into account an effective contribution of the
rapidly varying force to the average nonlinear dynamics and this
contribution might become large for $\delta = {\cal O}(1)$, i.e. when
$\epsilon \sim \omega^2$. Thus, the dynamics of the SG model with a rapidly
varying direct driving force may be described by a {\em renormalized}  SG
equation (\ref{18}) up to the terms of order of $\epsilon/\omega^2$.

The results obtained above may immediately be applied to describe the
renormalized dynamics of kinks in the presence of the rapidly varying ac
force. In fact, Eq. (\ref{18}) is the dc driven damped SG equation with a
{\em renormalized} coefficient in front of the term $\sim \sin \Phi$. This
simply means that we can apply all the results known for the standard SG
equation (see, e.g., \cite{McS,KM}) making only a {\em renormalization} of
the kink's width. For example, the kink solution of Eq.  (\ref{18}) at
$\gamma = f = 0$ has the form
\begin{equation}
\label{A1}
\Phi (x,t) = 4 \sigma \tan^{-1} \exp \left[ \frac{x - Vt}{l_0 \sqrt{1-V^2}}
\right],
\end{equation}
where $\sigma = \pm 1$ is the kink's polarity and $l_0 =
[J_0(\epsilon/\omega^2)]^{-1/2}$ is the kink's width at rest. The motion of
the kink in the presence of small dc force $f$ and damping $(\sim \gamma)$
is characterized by the steady-state velocity
\begin{equation}
\label{A2}
V_{*} = - \frac{\sigma}{\sqrt{1 + g^2}}, \;\;\;\;\; g \equiv
\left(\frac{4\gamma}{\pi f} \right) J_0\left(\frac{\epsilon}{\omega^2}
\right).
\end{equation}
In the theory of long Josephson junctions the kink's velocity is connected
with the voltage across the junction, $<\phi_t>$, where $<\ldots>$ stands
for the averaging in time, so that the result (\ref{A2}) for the
steady-state kink velocity gives the so-called zero-field steps in the
current-voltage (IV) characteristics of a long junction.  As follows from
Eq. (\ref{A2}), the renormalization of the parameter $g$ leads to a change
of the kink's velocity $V_{*}(f)$ and this, therefore, changes the slopes
of the voltage steps by the effect of the ac driving force.

\section{Parametric driving force}

Let us consider now a parametric driving force applied to the SG system,
with the main purpose to demonstrate that such a case is {\em very
different} from that analysed above.  The qualitative difference between
the effects produced by direct and parametric (rapidly oscillating) forces
is the following: A sufficient change of the system ``averaged'' dynamics
due to a rapidly oscillating direct force may be observed for amplitudes
$\epsilon \sim \omega^2$ whereas in the case of a parametric force, similar
effects may be already observed for {\em smaller} amplitude, {\em i.e.}, in
fact for $\epsilon \sim \omega$.  To prove this statement and to show how
our asymptotic method works for the case of the parametric force, we
consider the parametrically perturbed SG equation in the form
\begin{equation}
\label{1s}
\frac{\partial^2 \phi}{\partial t^2} - \frac{\partial^2 \phi}{\partial x^2}
+ \sin \phi = f - \gamma \frac{\partial \phi}{\partial t} + \epsilon \sin
\phi \cos (\omega t),
\end{equation}
where $f$ and $\gamma$ have the same sense as above, but this time
$\epsilon$ is the amplitude of the parametric force. Various applications
of the model (\ref{1s}) were discussed in the review paper \cite{KM} (see
also Refs. \cite{PSG}).

We assume that the parametric force is rapidly oscillating, {\em i.e.}, the
frequency $\omega$ is large.  As above, we look for a solution of Eq.
(\ref{1s}) in the form of asymptotic expansion
\begin{equation}
\label{B1}
\phi = \Phi + A \cos(\omega t) + B \sin (\omega t) + C \cos (2\omega t) + D
\sin (2\omega t) + \ldots,
\end{equation}
where the functions $\Phi$, $A$, $B, \ldots$ are assumed to be slowly
varying on the time scale $\sim \omega^{-1}$. The function $\Phi$ in Eq.
(\ref{B1}) determines, in fact, the evolution of the averaged field
component because $<\phi> = \Phi$, where the brackets $<\ldots>$ stand for
the averaging in fast oscillations.  Substituting the expression (\ref{B1})
into Eq. (\ref{1s}) and collecting, as in the case of the direct driving
force,  all the coefficients in front of the different harmonics, we again
obtain an infinite set of coupled nonlinear equations.  The subsequent (and
very important) step of such an analysis is to find the form of the
asymptotic expansions for the coefficients $A$, $B, \ldots$. In the present
case it is easy to check that the expansions (\ref{11}) do not give a
closed asymptotic procedure, and in the case $\epsilon \sim \omega^{2}$ the
driving force from Eq.  (\ref{1s}) contributes to all the harmonics, so
that contributions of the other harmonics become large as well.  Comparing,
as in the previous case, different terms of the equations for the
coefficients $A, B, \ldots$, we may easily check that the asymptotic
procedure may be effectively formulated for {\em smaller} (but not small)
amplitudes, {\em i.e.}, when $\epsilon \sim \omega$, and, as above, it
gives all the corrections to the averaged nonlinear dynamics in a rigorous
way. Thus, we take the asymptotic expansions in the form
\begin{equation}
\label{11s}
A=\frac{a_{1}}{\omega^2}+\frac{a_{2}}{\omega^4}+ \ldots, \;\;
B=\frac{b_{1}}{\omega^3}+\frac{b_{2}}{\omega^5} + \ldots, \;\; C =
\frac{c_{1}}{\omega^4} + \ldots, \;\; D=\frac{d_{1}}{\omega^5}+ \ldots \;.
\end{equation}
Using the power series (\ref{11s}), we obtain the ``averaged'' equation in
the form
\begin{equation}
\label{12s}
\frac{\partial^2 \Phi}{\partial t^2} - \frac{\partial^2 \Phi}{\partial x^2}
+ \left( 1 - \frac{1}{4} \frac{a_1^2}{\omega^4} + \ldots \right) \sin \Phi
= f - \gamma \frac{\partial \Phi}{\partial t} + \frac{\epsilon}{2} \cos
\Phi \left( \frac{a_1}{\omega^2} + \frac{a_2}{\omega^4} + \ldots \right),
\end{equation}
The expansions (\ref{11s}) allow to find the coefficients of the asymptotic
expansions in each order in the small parameter $\omega^{-1}$, and all the
corrections are determined, as above, by {\em algebraic} relations.

In the first-order approximation only the term $\sim \epsilon a_1$
contributes to Eq. (\ref{12s}). From the asymptotic expansions it follows
that
\begin{equation}
\label{13s}
a_1 = - \epsilon \sin \Phi,
\end{equation}
and Eq. (\ref{12s}) yields
\begin{equation}
\label{14s}
\frac{\partial^2 \Phi}{\partial t^2} - \frac{\partial^2 \Phi}{\partial x^2}
+ \left( 1 + \frac{1}{2} \Delta^2 \cos \Phi \right) \sin \Phi = f - \gamma
\frac{\partial \Phi}{\partial t},
\end{equation}
where $\Delta \equiv \epsilon /\omega$. Equation (\ref{14s}) takes into
account an effective contribution of the rapidly varying parametric force
to the slowly varying nonlinear dynamics {\em in the lowest order}, and all
the corrections coming from the approximation of the next order are
proportional to the small parameter $\sim \omega^{-2}$.  However, even the
lowest-order contribution might become large for $\Delta = {\cal O}(1)$,
{\em i.e.}, when $\epsilon \sim \omega$.

Thus, the ``averaged'' dynamics of the SG model with a rapidly varying
parametric force is described by the double SG equation (\ref{14s}). As a
matter of fact, the double SG equation is rather well studied (see, e.g.,
Refs. \cite{Con,Sod} and references therein) and properties of its kink
solutions are known as well. In particular, the kink solution of Eq.
(\ref{14s}) at $f = \gamma = 0$ may be written in the form \cite{Con}
\begin{equation}
\label{R1}
\Phi(x,t) = 2 \tan^{-1} \left[ \frac{1}{\sqrt{1 + \Delta^2/2}} {\rm cosech}
\left( \sqrt{1 + \frac{\Delta^2}{2}} \frac{x - Vt}{\sqrt{1 - V^2}} \right)
\right],
\end{equation}
and this solution may be treated as two coupled $\pi$-kinks. In Fig. 1 we
show the results of numerical simulations of the parametrically driven SG
system, Eq. (23). In all the cases analysed in the present paper we have
integrated the driven damped SG equation on a spatial interval of length
$L$, with periodic boundary conditions. As is seen from Fig. 1, the
sech-type shape of the $2\pi$ kink corresponding to the standard
(unperturbed) SG system is modified, and the function $\phi_x$ displays a
two-peaked profile which, as a matter of fact, is one of the main features
of the kink solution (\ref{R1}). Increasing $\Delta^2$ one may observe, in
accordance with Eq. (\ref{R1}), that the function $\phi$ has an evident
shape of two $\pi$-kinks separated by a distance $\sim \Delta^2$.

As has been shown in Ref. \cite{3}, $\pi-$kinks themselves may exist in the
parametrically driven SG chain provided the condition $\Delta^2 >2$ is
satisfied. This condition simply means that the effective ``averaged''
potential for the slowly varying field component $\Phi$ exhibits a local
minimum at $\Phi = \pi$ so that this stationary state becomes stable.

The appearence of new features in the slowly varying (``averaged'') system
dynamics of the SG system for $\Delta^2 >2$ is similar to the phenomenon of
the parametric stabilization of the reverse pendulum in the well known
Kapitza problem \cite{1,N}.  However, in the problem under consideration
some novel features in the nonlinear dynamics of the parametrically driven
SG system may be really observed for {\em any value of the effective
parameter} $\Delta^2$.  Indeed, as is known from the theory of the double
SG equation \cite{Sod}, at any value of $\Delta^2$ the kink (\ref{R1})
possesses the so-called internal (``shape'') mode which describes
variations of the kink's width. This internal mode is absent for the
standard SG kink, and the mode frequency $\Omega^2_{sol}$ splits at any
$\Delta^2 \neq 0$ from the gap frequency of the linear spectrum. For
$\Delta^2 >2$ the kink's internal mode may be described as relative
oscillations of the $\pi-$ kinks of which the $2\pi-$kink consists.
However, this mode does, in fact, exist at any value of the effective
parameter $\Delta^2$, and it may be observed as periodic variations of the
kink's shape.

We have measured numerically the frequency of the shape oscillations of the
$2\pi$-kink (\ref{R1}) directly solving Eq. (\ref{1s}) and also using the
averaged equation (\ref{14s}). The numerical results are shown in Fig. 2
for selected values of the external frequency, $\omega = 50$ and $\omega =
100$. For relatively small $\Delta^2$ ({\em i.e.}, $\epsilon$ in Fig. 2),
when higher-order corrections to Eq. (\ref{14s}) are negligible, a perfect
agreement between the results for the parametrically driven SG model
(\ref{1s}) and those for the averaged equation (\ref{14s}) are clearly
observed, justifying the validity of our asymptotic procedure.

\section{Kinks on rotating and oscillating backgrounds}

As was mentioned in Ref. \cite{6}, the other physically important case when
a rapidly varying ac force may change drastically the kink dynamics is the
case of a rotating and oscillating background. We should note, however,
that if one considers relatively small system lengths $L$, even a
relatively weak driving force may lead to complicated dynamics involving
coexisting states of bunched kinks and nontrivial background states
\cite{Rot}. These latter effects are probably caused by the influence of
nonzero boundary conditions which may ``help to lock'' kink-like states on
rotating backgrounds. Here we are interested in the dynamics of the long SG
systems (i.e. the kink's length is much smaller than the system length $L$)
when the high-frequency force phase-locks the SG field in an oscillating
and rotating state and thereby creates a mechanizm (an effective
gravitation field) for supporting kink solitons. However, we should note
that the theory presented below can not be applied to formally infinite SG
system because for the case of the continum linear spectrum the applied ac
force may create resonances making the system dynamics much more
complicated and even chaotic. In fact, we need finite-width (but of large
$L$) systems in order to avoid linear resonances if the frequency of the
external ac force is selected in a gap between the nearest
eigenfrequencies.

To describe the effect of the kink phase-locking on a rotating background
analytically in a rigorous way, we consider the perturbed SG equation
(\ref{1}) assuming $f>1$, in which case the ground state of the SG chain is
not stable and the chain rotates with the frequency $\Omega$ so that $\phi
\approx \Omega t$.  Applying the high-frequency ac force $\sim \epsilon$ we
are interested in the slowly varying phase-locked system dynamics on such a
rotating background.  Accordingly, we look for a solution of Eq. (\ref{1})
in the form
\begin{equation}
\label{c2}
\phi = \Phi + \Omega t + \xi,
\end{equation}
where $\xi$ is the rapidly varying part oscillating with the large
frequency $\omega$ of the external ac force, $\Phi$ is the slowly varying
(long time scale) part, and $\Omega$ is the average frequency of rotation
for the background field, which we assume to be phase-locked to the
external ac field, i.e.  $\Omega = \pm k \omega$, $k$ being integer.
Looking for the rapidly oscillating part $\xi$ in the form of a Fourier
series with slowly varying coefficients,
\begin{equation}
\xi = A \cos(\omega t) + B \sin (\omega t) + \ldots,
\end{equation}
we obtain the following equations for the averaged field component $\Phi$
and the expansion coefficients $A, B, \ldots$,
\begin{equation}
\label{c3}
\frac{\partial^2 \Phi}{\partial t^2} - \frac{\partial^2 \Phi}{\partial x^2}
+ J_k(A) \sin \Phi = f - \gamma k \omega - \gamma \frac{\partial
\Phi}{\partial t},
\end{equation}
\begin{equation}
\label{c4}
\left(- \omega^2A - 2\omega \frac{\partial B}{\partial t} +
\frac{\partial^2 A}{\partial t^2} \right) - \frac{\partial^2 A}{\partial
x^2} + [J_{k+1}(A) - J_{k-1}(A)] \cos \Phi + \gamma \left( \frac{\partial
A}{\partial t} + \omega B \right) = \epsilon,
\end{equation}
\begin{equation}
\label{c5}
\left(- \omega^2B + 2\omega \frac{\partial A}{\partial t} +
\frac{\partial^2 B}{\partial x^2}\right) - \frac{\partial^2 B}{\partial
x^2} - [J_{k+1}(A) + J_{k-1}(A)] \sin \Phi + \gamma \left( \frac{\partial
B}{\partial t} - \omega A\right) = 0,
\end{equation}
and so on. Unlike the case considered in Sec. II, in the present problem
there are two rapidly oscillating contributions with the frequencies
$\omega$ and $k \omega$, so that the final equations (\ref{c3}) to
(\ref{c5}) for the slowly varying coefficients differ from the
corresponding equations (\ref{4}) to (\ref{7}) of Sec. II. To take into
account the second term in the right-hand side of Eq.  (\ref{c3}) in a
self-consistent way, we also assume the dissipation to be rather small,
$\gamma \omega \sim 1$, which is a typical case for Josephson junctions.

Making now asymptotic expansions similar to the case of the direct ac force
considered above, {\em i.e.},
\begin{equation}
\label{c6}
A = a_1 + \frac{a_2}{\omega^2} + \ldots, \;\; B = \frac{b_1}{\omega} +
\ldots,
\end{equation}
we find the following relations [{\em cf.} Eqs. (\ref{12}), (\ref{14})]
\begin{equation}
\label{c7}
a_1 = - \frac{\epsilon}{\omega^2} \equiv - \delta, \;\;\; b_1 = - \gamma
a_1,
\end{equation}
which allow to obtain the effective equation for the slowly varying system
dynamics by just combining Eqs. (\ref{c7}), (\ref{c6}) and (\ref{c3}). The
final equation is just Eq. (\ref{c3}) with $A= - \epsilon/\omega^2$, which
describes the kink dynamics on the background rotating with the frequency
$\Omega = \pm k \omega$. It is important to note that the resulting
(effective) dc force in the ``averaged'' nonlinear equation (\ref{c3}) is
represented by the term $f - \gamma k \omega$ but not $f$ itself, {\em
i.e.}, the kink on the rotating and oscillating background may move even in
the absence of the constant contribution to the bias current, $f=0$. Figure
3 shows the results of the numerical calculation of the first Shapiro step
($k = 1$) of a long Josephson junction described by the model (\ref{1})
with the parameters $\gamma =0.1$, $\epsilon = 12.5$, and $\Omega = \omega
= 2.5$.  As is clearly seen from that figure, the steps cross the zero
current axis, displaying the property mentioned above; constant-voltage
zero-crossing steps are of considerable practical interest for
voltage-standard applications of Josephson junctions (see, {\em e.g.},
\cite{FMP} and references therein).  If we select $f=0.25$, the effective
force acting on the kink vanishes, and the kink is observed at rest (see
Fig. 4).  This result is in excellent agreement with the averaged SG
equation (\ref{c3}), where the effective force is found to be $f-\gamma k
\omega$. Increasing the value of the bias current $f$, we create an
effective force acting on the kink and it moves to the left (see Fig. 5).
It is very interesting to note that the kink motion is even possible in the
absence of the constant component of the bias current, {\em i.e.}, at
$f=0$. Figure 6 shows this case, when the effective force $-\gamma k
\omega$ generates the motion of the kink in the direction opposite to that
shown in Fig. 5.

As a matter of fact, our approach and the resulting equation for the slowly
varying field component $\Phi$ do not specify exactly the type of nonlinear
solutions we deal with. This means that the consideration described above
may be effectively applied to other problems, much more general than a
single-kink propagation. One of the important generalizations of this
approach is to treat multi-soliton (multi-kink) problems. In particular,
for the nonlinear dynamics on oscillating and rotating background the
effects similar to those described for a single kink may be also observed
for the case of two, three, and more kinks. In particular, Figures 7 and 8
present results of our numerical simulations of the same effects as for the
single kink, but for the case of two kinks in the directly driven SG model
(\ref{1}). As may be noted from these figures, the large-amplitude direct
ac force generates some radiation, especially for the cases where the kinks
are observed at rest, but still kinks exist as localized and well defined
objects.

\section{Conclusions}

In conclusion, we have analysed the dynamics of SG kinks in the presence of
rapidly varying periodic perturbations of different physical nature. We
have proposed a rigorous analytical approach to derive the ``averaged''
equation for the slowly varying field component, and we have demonstrated
that in the main order of the asymptotic procedure the effective equation
is a renormalized SG equation in the case of the direct driving force or
rotating (and phase-locked to the external ac driving force) background,
and it is a double SG equation for the parametric driving force.  However,
the method itself does allow to find in a rigorous way the effective
equation for the slowly varying field component in any order of the
asymptotic expansion in the parameter $\omega^{-1}$, $\omega$ being the
frequency of the rapidly varying perturbations which has been assumed to be
large.

Our main purpose was to show which kinds of qualitatively new physical
effects may be expected in dealing with the renormalized nonlinear dynamics
instead of unrenormalized one. In particular, we have predicted that the
parametric driving force may support oscillations of the kink's shape
(absent in the SG model)  which may be viewed as creation of a shape mode
of the $2\pi-$kink characterized by the internal frequency $\Omega_{sol}$.
For the problem of the kink propagation on rotating and oscillating
background, we have shown that a periodic ac force produces a drift in the
kink motion, which may be understood as an effect described by an effective
dc force to the kink motion in the fremework of an ``averaged'' nonlinear
dynamics.

One of the main conclusions of our analysis and numerical simulations, {\em
i.e.}, that the ``averaged'' nonlinear dynamics is drastically modified by
rapidly varying (direct or parametric) driving force but still may be
effectively described by renormalized nonlinear equations, is rather
general and applicable to many other nonlinear models supporting various
kinds of solitons.

\acknowledgments

YuK acknowledges hospitality of the Dipartimento di Fisica
dell'Universit\`{a} di Salerno during his short stay there, and also useful
discussions with A.  S\'anchez (Madrid) and K.H. Spatschek (D\"usseldorf).
His work was partially supported by the Australian Photonics Cooperative
Research Center (APCRC). Parts of this work were performed under the auspices
of the US Department of Energy. RDP acknowledges financial support from the
EC under contract no. SC1-CT91-0760 (TSTS) of the ``Science'' program, from
MURST (Italy), and from the Progetto Finalizzato ``Tecnologie
Superconduttive e Criogeniche'' del CNR (Italy).

\begin{figure}
\caption{The steady-state profile of the $\phi$ (a) and $\phi_x$ (b) fields
as a function of space in the case of a parametric driving force. The
parameters are $\gamma = 0.2$, $L = 10$, $\omega = 100$, and $\epsilon =
200$.}
\end{figure}

\begin{figure}
\caption{The frequency of the internal oscillations of the $2\pi$-kink,
$\Omega_{sol}$, as a function of the amplitude of the ac parametric force
$\epsilon$ for two values of the external force frequency, $\omega = 50$
(squares and crosses) and $\omega = 100$ (diamonds and plusses). The
results presented by diamonds and squares are obtained using the effective
double SG equation (\protect\ref{14s}) whereas the plusses and crosses are
the result of direct integration of the parametrically driven SG equation
(\protect\ref{1s}). Note that the agreement between the parametrically
driven model (\protect\ref{1s}) and the effective (``averaged'') model
(\protect\ref{14s}) is better for smaller $\Delta = \epsilon/\omega$, when
corrections to Eq. (\protect\ref{14s}) from the higher-order terms are
negligible.}
\end{figure}

\begin{figure}
\caption{The normalized IV curves (for the first Shapiro step)
characterizing the SG dynamics with the parameters $\gamma =0.1$, $L = 16$,
$\epsilon = 12.5$, and $\Omega = 2.5$; and with periodic boundary
conditions. Shown are the zero-field steps at $n= 0, 1, 2, 3, 4$, and $5$,
$n$ being the number of kinks in the system. Note that the IV curves
clearly cross the zero current axis and the steps are slightly asymmetric
around the voltage $<\phi_t> = 2.5$; the latter effect is caused by the
background oscillations and relatively small velocity.}
\end{figure}

\begin{figure}
\caption{The steady-state profile of the $\phi$ (a) and $\phi_{x}$ (b)
fields as a function of space over 10 periods of the external driving
force. Parameters are $L=16$, $\gamma =0.1$, $\epsilon = 12.5$, $\omega
=2.5$, and $f =0.25$. The value of $f$ is selected at the center of the
first Shapiro step and, therefore, the kink does not move as follows from
the theory because the effective force acting on a kink, $f-\gamma k\omega$,
is zero.}
\end{figure}

\begin{figure}
\caption{The same as in Fig. 4 but at $f=0.5$.}
\end{figure}

\begin{figure}
\caption{The same as in Fig. 4 but at $f=0$. The kink is moving due to the
uncompensated contribution of dissipative losses $ -\gamma k \omega $.}
\end{figure}

\begin{figure}
\caption{The same as in Fig. 4 but for the case of two kinks introduced by
a change of the boundary conditions.}
\end{figure}

\begin{figure}
\caption{The same as in Fig. 5 but for the case of two kinks.}
\end{figure}

\newpage
\begin{references}
\bibitem{1} L.D. Landau and M. Lifshitz, {\em Mechanics} (Pergamon Press,
Oxford, 1960).

\bibitem{N} J.A. Blackburn, H.J.T. Smith, and N. Gr{\o}nbech-Jensen, Am. J.
Phys. {\bf 60}, 903 (1992).

\bibitem{2} J. Mills, Phys. Lett. A {\bf 133}, 295 (1988); in particular,
this work has rediscovered the so-called zero Shapiro step, see S. Shapiro,
Phys. Rev. Lett. {\bf 11}, 80 (1963).

\bibitem{3}  Yu. S. Kivshar, N. Gr{\o}nbech-Jensen, and M.R. Samuelsen,
Phys. Rev. B {\bf 43}, 5698 (1992).

\bibitem{4} Yu.S. Kivshar, A. S\'{a}nchez, and L. V\'{a}zquez. Phys. Rev. A
{\bf 45}, 1207 (1992).

\bibitem{5} N. Gr{\o}nbech-Jensen and Yu.S. Kivshar, Phys. Lett. A {\bf
171}, 338 (1992)

\bibitem{6} N. Gr{\o}nbech-Jensen, Yu.S. Kivshar, and M. Salerno, Phys.
Rev. Lett. {\bf 70}, 3181 (1993).

\bibitem{NL} N. Gr{\o}nbech-Jensen and P. Lomdahl, Phys. Lett. A {\bf 177},
351 (1993).

\bibitem{McS} D.W. McLaughlin and A.C. Scott, Phys. Rev. A {\bf 18}, 1652
(1978).

\bibitem{KM} Yu.S. Kivshar and B.A. Malomed, Rev. Mod. Phys. {\bf 61}, 763
(1989).

\bibitem{PSG} N. Gr{\o}nbech-Jensen, Yu.S. Kivshar, and M.R. Samuelsen,
Phys. Rev. B {\bf 43}, 5398 (1991); R. Grauer and Yu.S. Kivshar, Phys. Rev.
E {\bf 48} (1993) [EF 5050].

\bibitem{Con} C.A. Condat, R.A. Guyer, and M.D. Miller, Phys. Rev. B {\bf
27}, 474 (1983).

\bibitem{Sod} D.K. Campbell, M. Peyrard, and P. Sodano, Physica D {\bf 19},
165 (1986).

\bibitem{Rot} G. Rotoli, G. Costabile, and R.D. Parmentier, Phys. Rev. B
{\bf 41}, 1958 (1990).

\bibitem{FMP} G. Filatrella, B.A. Malomed, and R.D. Parmentier, Phys. Lett.
A {\bf 180}, 346 (1993).
\end{references}
\end{document}